\documentclass[doublecol]{epl2} 

\usepackage{graphics}
\usepackage{graphicx}

\title{Swimming active droplet: A theoretical analysis}
\shorttitle{Swimming active droplet}

\author{M. Schmitt \and H. Stark}
\shortauthor{M. Schmitt \etal}

\institute{                    
  Institut f{\"u}r Theoretische Physik, Technische Universit{\"a}t Berlin - Hardenbergstra{\ss}e 36, 10623 Berlin, Germany, EU
}


\abstract{
Recently, an active microswimmer was constructed where a micron-sized droplet of bromine water was placed into a surfactant-laden oil phase. Due to a bromination reaction of the surfactant at the interface, the surface tension locally increases and becomes non-uniform. This drives a Marangoni flow which propels the squirming droplet forward.
We develop a diffusion-advection-reaction equation for the order parameter of the surfactant mixture at the droplet
interface using a mixing free energy. Numerical solutions reveal a stable swimming regime above a
critical Marangoni number $M$ but also stopping and oscillating states when $M$ is increased further. The swimming
droplet is identified as a pusher whereas in the oscillating state it oscillates between being a puller and a pusher.
}
\begin{document}

\maketitle

\section{Introduction}
A rigorous understanding of swimming on the micron scale is crucial for developing microfluidic devices such as a lab-on-a-chip \cite{abgrall2007}. This understanding comes from watching nature, i.e., by studying the locomotion of living organisms such as bacteria or algae \cite{lauga2009} but also from designing artificial microswimmers, used for example as medical microrobots
\cite{nelson2010}. Both, real live cells and man-made microswimmers, have thoroughly been used to study interaction between swimmers \cite{ishikawa2007}, 
interaction with walls \cite{berke2008,drescher2011,zoettl2012,reddig2011}, or swarming \cite{lauga2012}. One possible design of an artificial swimmer is an active droplet. Here, we think of a droplet with a surface where a chemical reaction occurs. Alternatively, droplets or bubbles can be made active by having an internal heat source \cite{rednikov1994c}. Droplets are particularly interesting systems to study since they are used extensively in microfluidic devices as microreactors in which chemical or biological reactions take place \cite{teh2008, seemann2012}. In the following we give an example of an active droplet and investigate in detail its propulsion mechanism. 
Self-propelled active droplets have been studied in various experiments, including droplets on interfaces \cite{chen2009, bliznyuk2011}, droplets coupled to a chemical wave \cite{kitahata2011}, and droplets in a bulk fluid \cite{toyota2009, hanczyc2007, thutupalli2011, banno2012}. Theoretical treatments include a model of droplet motion in a chemically reacting fluid \cite{yabunaka2012}, studies of the stability of a resting droplet \cite{rednikov1994,rednikov1994b, velarde1996,velarde1998,yoshinaga2012}, and simulations of contractile droplets \cite{tjhung2012} and of droplets driven by nonlinear chemical kinetics \cite{furtado2008}. 

The swimming active droplet we consider in the following is a solution of water and bromine which is placed in a 
surfactant-rich oil phase \cite{thutupalli2011}. The resulting water droplet has a typical radius of $80\mu \mathrm{m}$. 
In order to lower the surface tension and thus the total energy of the system, the surfactants in the oil phase form a dense monolayer at the droplet interface, giving the droplet the structure of an inverse micelle. The observed directed swimming motion of the droplet with a typical swimming speed of $15\mu \mathrm{m/s}$ 
can be understood as follows \cite{thutupalli2011}. 

The bromine within the droplet chemically reacts with the surfactants in the interface which results
in a weaker surfactant. Hence, the 'bromination' reaction locally leads to a higher surface tension in the interface. As a consequence local gradients in surface tension will lead to a fluid flow at the interface and in the adjacent fluid inside and outside of the droplet in the direction of increasing surface tension. This effect is called Marangoni effect. 
The fluid flow then leads in turn to advection of surfactants at the interface. As a result gradients in surface tension are enhanced. Thus, the resting state becomes unstable and the droplet starts to move. Additionally, brominated surfactants are constantly replaced by non-brominated surfactants from the oil phase by means of desorption and adsorption. The droplet stops to swim when either the bromine or the non-brominated surfactants in the oil phase are exhausted. This was also observed in the experiments \cite{thutupalli2011}. 

The active droplet is an interesting realization of the 'squirmer' \cite{lighthill1952,blake1971} which has been introduced to model the locomotion of microorganisms. Often they propel themselves by a carpet of beating short filaments called cilia on their surfaces. Instead of modeling each cilium separately, one prescribes the fluid flow at the surface initiated by the beating cilia which then drags the squirmer through the fluid. Here, for the active droplet the surface flow is generated by the Marangoni effect.

The swimming active droplet crucially depends on the dynamics of the mixture of non-brominated and brominated surfactants at the interface. 
In this article we model it by means of a diffusion-advection-reaction equation based on a free energy functional for the surfactant mixture. Numerical solutions then show that in a certain parameter range the resting state of the droplet becomes unstable and the droplet starts to move. The solutions reach a stationary state corresponding to a swimming motion and confirm that the droplet is a pusher \cite{lauga2009}, as found in the experiments\ \cite{thutupalli2011}. In addition, we identify further patterns of motion. We find that the droplet stops after an initial motion or that it oscillates back and forth.

\section{Model} 

In order to model the droplet propulsion we set up a dynamic equation for the surfactant mixture 
at the droplet interface that includes all processes mentioned before. We assume that the surfactant completely covers the droplet interface
without any intervening solvent. We also assume that the area of both types of surfactant molecules (brominated and non-brominated) is the same. Denoting the brominated surfactant density by $c_1$ and the non-brominated density by $c_2$, we can therefore set $c_1+c_2=1$. We then take the concentration difference between brominated and non-brominated surfactants as an order parameter $\phi=c_1-c_2$.
In other words $\phi=1$ corresponds to fully brominated and $\phi=-1$ to fully non-brominated and $c_1=(1+\phi)/2$ and $c_2=(1-\phi)/2$. Finally, we choose a constant droplet radius $R$.

\subsection{Diffusion-Advection-Reaction equation}

The dynamics of $\phi$ is governed by a diffusion-advection-reaction equation:
\begin{equation}
   \dot{\phi}=-\nabla\cdot(\mathbf{j}_D+\mathbf{j}_A) - \tau_R^{-1} (\phi-\phi_{eq})\;,\label{eq:conti}
\end{equation}
with diffusive current $\mathbf{j}_D$ and advective current $\mathbf{j}_A$ due to the Marangoni effect. The
third term on the right-hand side of Eq.\  (\ref{eq:conti}) is the reaction term and describes the bromination reaction as well as desorption of brominated and adsorption of non-brominated surfactants to and from the outer fluid. $\tau_R$ is the timescale on which these processes happen and ${\phi_{eq}}$ sets the equilibrium coverage of $\phi$. In other words, ad- and desorption dominates for $\phi_{eq}<0$, while bromination dominates for $\phi_{eq}>0$. Imagine for instance the case $\phi_{eq}=1$, i.e., a droplet with bromination but without ad- and desorption
of surfactants. The reaction term would then always be positive, therefore driving the droplet to a completely brominated state $\phi=1$. 

The general mechanism of Eq.\ (\ref{eq:conti}) is as follows. The diffusive current always points 'downhill', 
$\mathbf{j}_D\propto -\nabla\phi$. However, we will show below that the opposite is true for $\mathbf{j}_A$ 
since approximately $\mathbf{j}_A\propto\nabla\phi$. Thus, apart from the reaction term, $\mathbf{j}_D$ and $\mathbf{j}_A$ are competing and as soon as $\mathbf{j}_A$ dominates over $\mathbf{j}_D$, $\phi$ experiences 'uphill' diffusion, i.e. phase separation. As a result the resting state will become instable and the droplet will start to move. 
We will now present a careful derivation of $\mathbf{j}_D$ and $\mathbf{j}_A$ from a free energy approach.
This shows that the diffusive and advective currents in Eq.\ (\ref{eq:conti}) are in general non-linear functions of $\phi$.

\subsection{Diffusive current}

The basis for the following is a free energy density $f$ for the droplet interface, which we write down as a function of concentrations $c_1$ and $c_2$. In formulating the free energy density $f$, we follow the
Flory-Huggins approach \cite{flory1942}. Accordingly, $f$ is composed of the mixing entropy plus terms mimicking lateral attractive interaction between surfactants:
\begin{equation}
 f=\frac{k_B T}{A} \left[ c_1 \ln c_1+c_2 \ln c_2 -b_1 c_1^2-b_2c_2^2-b_{12} c_1c_2\right]\;,\label{eq:f}
\end{equation}
where $A$ denotes the area of a surfactant in the interface and $b_1$ ($b_2$) is a dimensionless parameter characterizing the interaction between brominated (non-brominated) surfactants and $b_{12}$ the interaction between different kind of surfactants. In the following we assume for simplicity $b_{12}=(b_1+b_2)/2$. 
In terms of the order parameter $\phi$ we obtain:
\begin{equation}
\left. \begin{array}{rl}
f(\phi)=\frac{k_B T}{A} \left[ \frac{1+\phi} 2 \ln \frac{1+\phi} 2+\frac{1-\phi} 2 \ln \frac{1-\phi} 2\right.\\
\\
\left.-\frac 3 8 (b_1+b_2)- \frac {\phi} 2 (b_1-b_2)-\frac{\phi^2} 8 (b_1+b_2) \right] .\end{array} \right.
\end{equation}
The total free energy is then given by the functional
\begin{equation}
 F[\phi]=\int f(\phi)\, dA\;.\label{eq:functional}
\end{equation}

For a conserved order parameter field the diffusive current is proportional to the gradient of the variation in $F$ with respect to $\phi$ \cite{bray2002}:
\begin{eqnarray}
   \mathbf{j}_D&=&-\lambda\nabla \frac{\delta F}{\delta \phi}=-\lambda f''(\phi) \nabla\phi\\
&=&-\frac{\lambda k_B T}{A} \left[\frac 1 {1-\phi^2}-\frac 1 4 (b_1+b_2)\right]\nabla\phi\;, \label{eq:j_d}
\end{eqnarray}
with positive mobility $\lambda$. Substituting $\mathbf{j}_D$ into Eq.\  (\ref{eq:conti}) yields a Cahn-Hilliard type equation \cite{desai2009}. 
Note that the diffusion constant in Eq.\ (\ref{eq:j_d}) decreases with increasing interaction energy. 
In fact, the condition $\mathbf{j}_D\propto-\nabla\phi$ is only fulfilled for a convex free energy with $f''(\phi)>0$, i.e. if $b_1+b_2<4$. In addition, the diffusion coefficient in $\mathbf{j}_D$ is smallest for $\phi = 0$. It increases
with $|\phi |$ and diverges at $|\phi | = 1$. An alternative approach of deriving diffusion currents in mixtures is presented in \cite{nauman2001, thiele2012}.

\subsection{Advective current}

\begin{figure}
\centering
\includegraphics[width=0.3\textwidth, angle=-90, trim=0mm 0mm 0mm 5mm, clip]{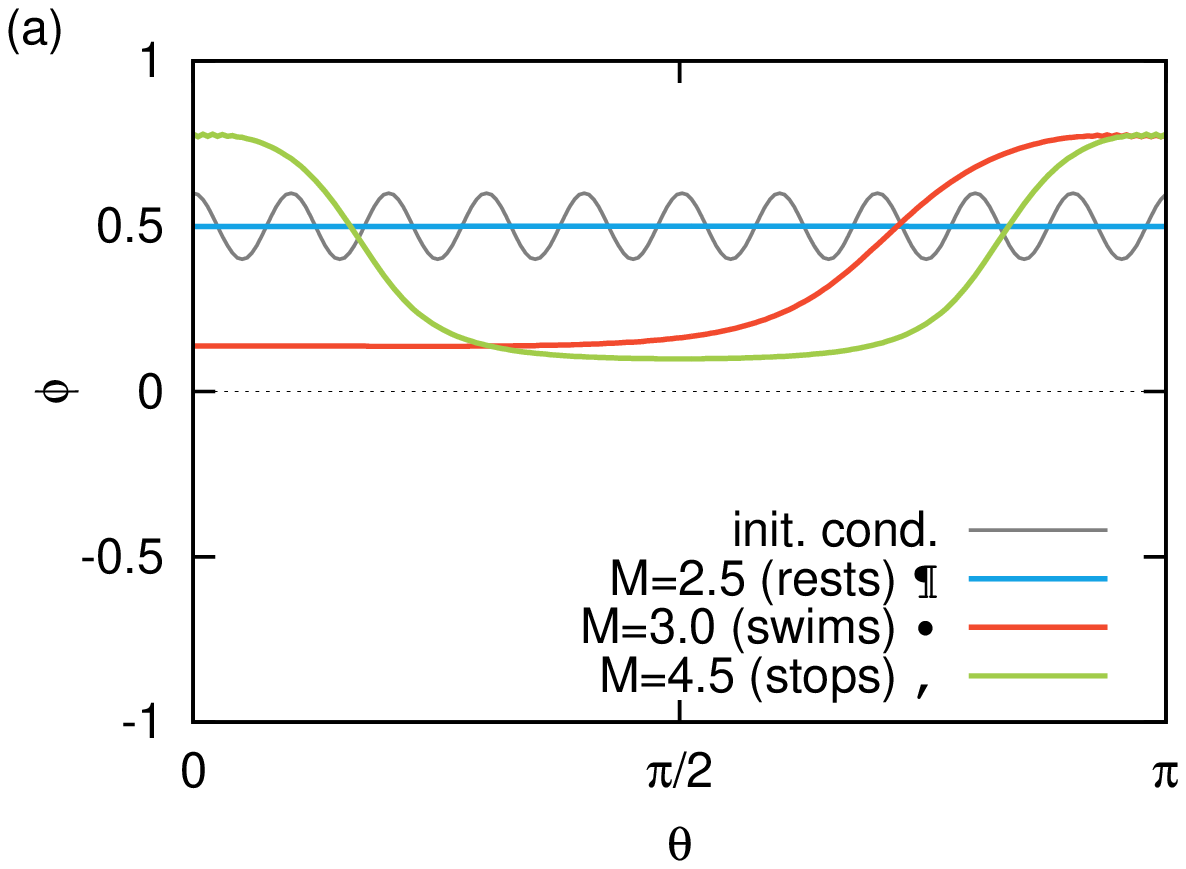}
\includegraphics[width=0.3\textwidth, angle=-90, trim=0mm 0mm 0mm 5mm, clip]{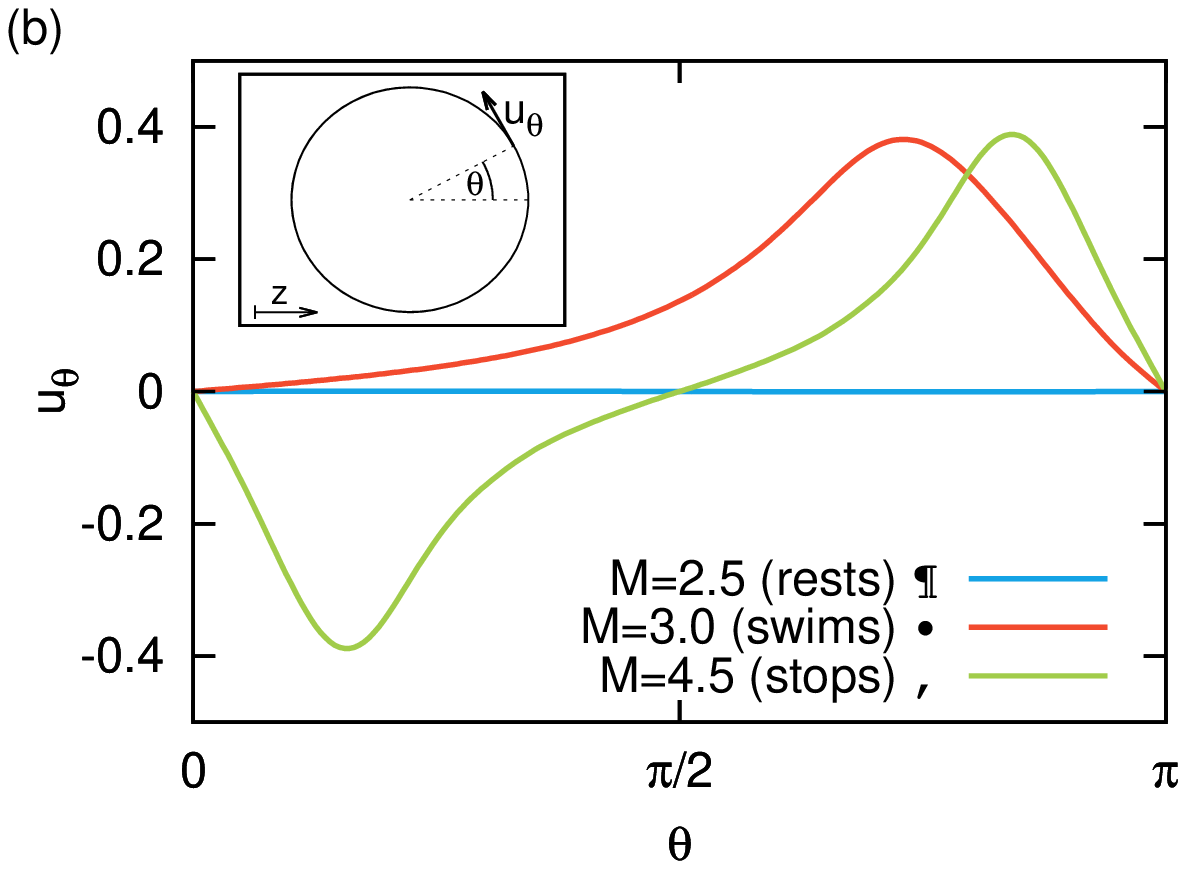}
\vspace{-0.2cm}
\caption{(a) Stationary order parameter profiles after $10^6$ time steps for $\phi_{eq}=0.5$ and several
Marangoni numbers $M$. Gray solid line: Initial condition. (b) Corresponding interface velocity profiles.
Inset: Droplet geometry.}
\label{fig:1}
\end{figure}

The advective current for the order parameter $\phi$
is given by
\begin{equation}
 \mathbf{j}_A=\phi \mathbf{u}\;,\label{eq:j_a}
\end{equation}
where $\mathbf{u}$ is the velocity of the surfactants at the interface.\footnote{Let the advective currents of the two types of surfactants be $\mathbf{j}^1_{A}=c_1 \mathbf{u}_1$ and $\mathbf{j}^2_{A}=c_2 \mathbf{u}_2$. Under the assumption that the individual velocities are identical $\mathbf{u}_1=\mathbf{u}_2=\mathbf{u}$, one obtains $\mathbf{j}_A=\mathbf{j}_A^1-\mathbf{j}_A^2 =\phi \mathbf{u}$.} Since we are studying the active droplet in an axisymmetric geometry, we assume $\phi=\phi(\theta)$ and $\mathbf{u} =u_{\theta}(\theta)\mathbf{e}_{\theta}$, where the front of the droplet is at $\theta=0$, see inset of Fig. \ref{fig:1} (b). For this geometry there exists a solution of the Stokes equation for the fluid flow field inside and outside of the droplet as well as the fluid velocity at the interface \cite{levan1976,mason2011}. The solution at the interface is given in terms of the surface tension gradient: 
\begin{equation}
\left. u_{\theta} \right |_{r=R} =\sum_{n=2}^{\infty} \frac {n(n-1)} {2\eta}\left[\int_0^{\pi} C_n^{-1/2}(z')\frac{d \sigma}{d\theta'}d\theta'\right] \frac{C_n^{-1/2}(z)}{\sin\theta}\;,\label{eq:u}
\end{equation}
where $z=\cos(\theta)$. $\eta=\eta_i+\eta_o$ is the sum of the viscosities inside and outside of the droplet and $C_n^{-1/2}$ are Gegenbauer polynomials. They are related to Legendre polynomials by 
$P_n(z)=-\frac{d}{dz}C_{n+1}^{-1/2}(z)$. 
Equation (\ref{eq:u}) is nothing but a representation of the the Marangoni effect. It essentially states $\mathbf{u}\propto \nabla\sigma$, i.e., a fluid flow in the direction of $\nabla\sigma$.

Thus, in order to calculate $u_{\theta}$, we need an expression for ${d \sigma}/{d\theta}$, which can be found by deriving an equation of state for the surface tension $\sigma$. The surface tension $\sigma$ is the thermodynamic force conjugate to the surface area. This gives:
\begin{equation}
 \sigma=f-\frac {\partial f}{\partial c_{1}} c_1-\frac {\partial f}{\partial c_{2}}c_2\;,
\end{equation}
which we identify as the Legendre transform of the free energy (\ref{eq:f}) to the chemical potentials $\mu_i=\frac{\partial f}{\partial c_i}$. Hence, ${\sigma=\frac{k_B T}{A}\left[ b_1c_1^2+b_2c_2^2+b_{12}c_1c_2\right],}$
or in terms of $\phi$ and again with $b_{12}=(b_1+b_2)/2$:
\begin{equation}
 \sigma(\phi)=\frac{k_B T}{4A}\left[ \frac 9 8 (b_1+b_2) + 2(b_1-b_2) \phi+\frac 7 8 (b_1+b_2) \phi^2 \right] \;.
\end{equation}
In order to obtain the proper behavior of the equation of state, i.e. an increasing surface tension with increasing $\phi$, we need to assure that $\sigma'(\phi)>0$. This holds if $b_1>b_2$, meaning that the interaction energy between brominated surfactants has to be higher than between the non-brominated ones. Note that in the limit of $\phi\rightarrow 0$ the equation of state becomes linear in $\phi$. The gradient of $\sigma$ is given by
\begin{equation}
 \frac{d \sigma}{d\theta}=\sigma'(\phi)\frac{d \phi}{d\theta}=\frac{k_B T}{2A}(b_1-b_2)\left[1+\frac 7 {8} \frac{b_1+b_2}{b_1-b_2}\phi\right] \frac{d \phi}{d\theta}\;.\label{eq:dsigma_dtheta}
\end{equation}
By substituting this into Eq.\ (\ref{eq:u}), one can calculate the advective current (\ref{eq:j_a}) for a given $\phi(\theta)$. 

Eqs. (\ref{eq:u}) and (\ref{eq:dsigma_dtheta}) essentially state that $\mathbf{u}\propto\nabla\phi$. Therefore,
when $\phi>0$, the advective current $\mathbf{j}_A=\phi\mathbf{u}$ apparently always points 'uphill', i.e., in
the opposite direction compared to $\mathbf{j}_D$.  On the other hand, when $\phi<0$, the advective current acts 'downhill', i.e., in the same direction as $\mathbf{j}_D$. As a consequence, the Marangoni flow will only drive the droplet when $\phi>0$. This is the case when there are more brominated surfactants than non-brominated ones. 

Together with (\ref{eq:j_d}) and (\ref{eq:j_a}), Eq.\ (\ref{eq:conti}) becomes a closed equation for $\phi$. 
Writing gradients in units of $R^{-1}$ and time in units of the diffusion time scale
$\tau_D=R^2A (\lambda k_B T)^{-1}$ yields
\begin{equation}
   \dot{\phi}=-\nabla\cdot(\mathbf{j}_D+M \phi \mathbf{u}) - \kappa (\phi-\phi_{eq})\;,\label{eq:conti2}
\end{equation}
where the currents $\mathbf{j}_D$ and $\mathbf{j}_A = M \phi \mathbf{u}$ are now dimensionless and
\begin{equation}
 M=\frac{ (b_1-b_2) R}{\lambda\eta}\;,
\end{equation}
is called Marangoni number. This number compares the advective current due to the Marangoni effect,
which scales as $k_B T (b_1-b_2) (R A\, \eta)^{-1}$,  to the diffusive current. Accordingly, $\kappa = \tau_D\tau_R^{-1}$ is the ratio between diffusion and reaction time scale. 

\section{Results}  

We numerically solve the diffusion-advection-reaction equation for $\phi$ with the initial condition 
$\phi(\theta)=\phi_{eq} + \delta\phi(\theta)$, where $\delta\phi(\theta)$ is a small perturbation 
[solid line in Fig.\ \ref{fig:1}(a)]. The boundary conditions at $\theta=0,\pi$ are given by a vanishing current, 
$\mathbf{j}_D+M\phi \mathbf{u}=0$. We keep $\kappa$ fixed to a value of $0.1$ for all numerical solutions and comment later on the impact of $\kappa$ on the results. Therefore,
we are left with the Marangoni number $M$ and $\phi_{eq}$ as the crucial parameters. To assure a convex free energy, we set $b_1+b_2=3$.

\subsection{Order parameter and velocity profiles} 

\begin{figure}
\centering
\includegraphics[width=0.6\textwidth, angle=-90, trim=0mm 8mm 0mm 120mm, clip]{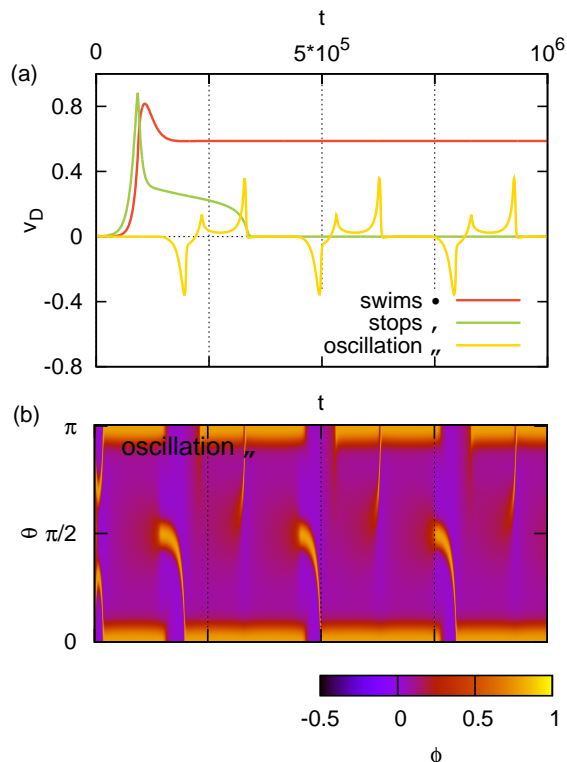}\\
\vspace{-1cm}
\caption{(a) Droplet swimming velocity $v_D$ for swimming, stopping, and oscillating droplets. Parameters are 
the same as in Fig.\ \ref{fig:1} and case 4 belongs to $M=10.5$. (b) Depiction of the chemical wave of case 4 in a $\phi(\theta,t)$ plot. Same timescale as in (a).
}
\label{fig:2}
\end{figure}

Figure\ \ref{fig:1}(a) shows examples of the stationary order parameter profile for $\phi_{eq}=0.5$ and different values of $M$ together with the corresponding interface velocity profiles in Fig.\ \ref{fig:1}(b). Starting with a small Marangoni number of $M=2.5$, the order parameter relaxes into the homogeneous trivial solution $\phi=\phi_{eq}$ of Eq.\ (\ref{eq:conti}), thus the droplet rests. Above a critical Marangoni number, the order parameter evolves to a stationary inhomogeneous profile, as Fig.\ \ref{fig:1} shows for $M=3$. 
In parallel, the droplet velocity $v_D$ depicted in Fig.\ \ref{fig:2}(a) saturates on a non-zero value. 
The droplet swimming speed is given by $v_D =(6\eta_i+4\eta_o)^{-1} \int_0^{\pi} \sin^2 \theta \frac{d\sigma} {d \theta} d\theta$ \cite{levan1976}. Since $C_2^{-1/2}(\cos(\theta))=\sin^2(\theta)/2$, $v_D$ is determined by the first coefficient of the sum in (\ref{eq:u}) and thus $v_D =\frac 8 {\pi} \frac {\eta_i+\eta_o} {6\eta_i+4\eta_o} \int_0^{\pi} \sin \theta u_{\theta} d\theta$. Note that in our approach $v_D$ reaches a stationary value without having to introduce a 'backward' Marangoni stress, as suggested in \cite{thutupalli2011}. Further increasing the Marangoni number to $M=4.5$, the droplet starts to swim but then stops rapidly. 
The stationary order parameter profile becomes symmetric around $\theta=\pi/2$ and swimming is
not possible. Finally, the droplet reaches an oscillating state for even higher Marangoni numbers where
it oscillates back and forth as the droplet swimming speed in Fig. \ref{fig:2}(a) demonstrates.
In this case the order parameter $\phi(\theta, t)$ resembles a chemical wave that travels back and forward between 
$\theta=0$ and $\theta=\pi$. The wave is depicted in Fig. \ref{fig:2}(b). Note that the frequency of the oscillation increases with $M$.
Finally, we remark that from comparing Figs.\ \ref{fig:1} (a) and (b), it is now apparent that indeed Eq.\ (\ref{eq:u}) essentially gives $\mathbf{u}\propto\nabla\phi$.

\subsection{Phase diagram}

\begin{figure}
\centering
\includegraphics[width=0.4\textwidth,trim=0mm 0mm 5mm 0mm]{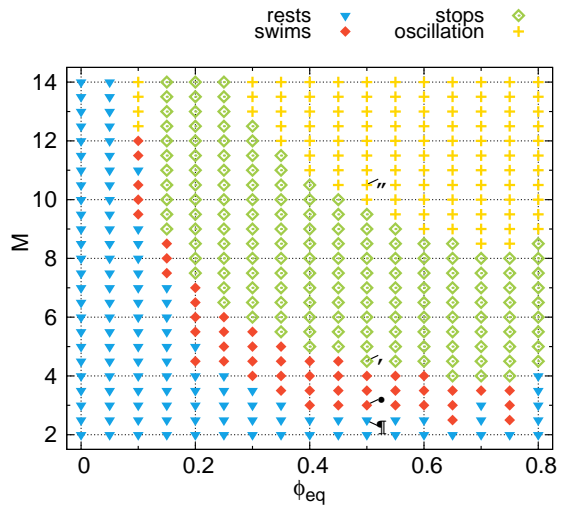}
\vspace{-0.5cm}
\caption{Phase diagram of the active droplet in ($\phi_{eq},M$) parameter space. Examples for the
order parameter profiles at the positions marked with numbers are given in Fig. \ref{fig:1}(a) (regime 1-3)
and Fig. \ref{fig:2}(b) (regime 4).}
\label{fig:3}
\end{figure}

Figure \ref{fig:3} shows the phase diagram in ($\phi_{eq},M$) parameter space with the four regimes
of the droplet dynamics: resting, swimming, stopping, and oscillating. Since there is no swimming motion possible for negative $\phi_{eq}$, as discussed before, we only
show the phase diagram in the range $0\leq\phi_{eq}\leq 0.8$.\footnote{Due to the $\phi$ dependent
diffusion coefficient in Eq.\ (\ref{eq:j_d}), numerics requires a much finer grid above $\phi_{eq}=0.8$. However, in several tests for different values of $M$ no swimming solutions were found above $\phi_{eq}=0.8$.} We find similar phase diagrams for smaller
values of $\kappa$. For $\kappa=0.01$ the swimming region increases in size and then shrinks
again for $\kappa=0.001$ until for $\kappa=0$ swimming solutions are no longer possible.
The critical Marangoni number at the onset of the swimming regime stays, however, roughly constant. 
On the other hand,  for $\kappa=1$ and $10$, i.e., in the limit of fast bromination reaction and exchange of surfactants, only resting, stopping and oscillating solutions but no stable swimming solutions were found.

\subsection{Reduced phase space}
\begin{figure}
\centering
\includegraphics[width=0.44\textwidth, angle=-90, trim=75mm 25mm 5mm 130mm, clip]{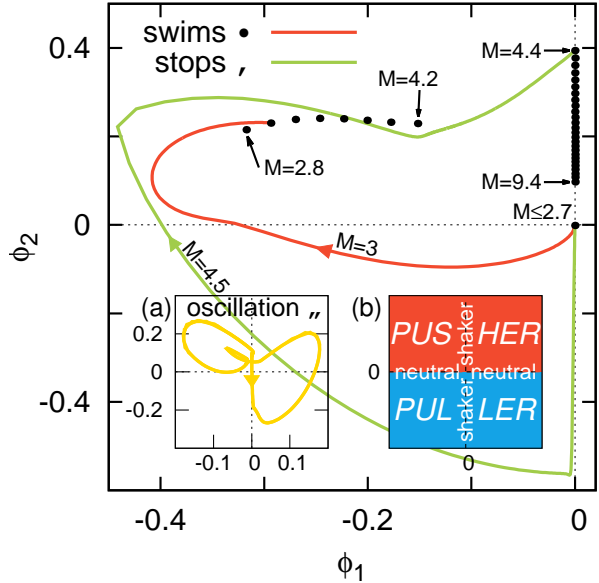}
\vspace{-0.2cm}
\caption{Droplet dynamics in the reduced phase space ($\phi_1,\phi_2$). Black dots show the fixed points
for different values of $M$; from dot to dot $M$ increases by $0.2$. The red and the green line show,
respectively, trajectories in the swimming ($M=3$) and stopping ($M=4.5$) state. Inset (a): limit cycle in the oscillating state ($M=10.5$). Inset (b): map for the swimmer type in ($\phi_1,\phi_2$) space classified by the 
stirring parameter $\beta=-\phi_2/|\phi_1|$ (see main text).}
\label{fig:4}
\end{figure}

Due to the axisymmetric geometry we decompose the order parameter $\phi$ into Legendre modes
\begin{equation}
\phi(\theta,t)=\sum_{n=0}^{\infty} P_n(\cos(\theta))\phi_n(t)\;.
\end{equation}
Together with Eqs.\ (\ref{eq:dsigma_dtheta}) and (\ref{eq:u}) one obtains an expression for $u_{\theta}$
 as a function of the mode amplitudes $\phi_n$. $\phi_1$ determines the swimming speed and $\phi_{n>1}$ 
 corresponds to the higher modes of $u_{\theta}$. In the following, we use the initial condition 
 $\phi_0(t=0)=\phi_{eq}$. In order to investigate the four  regimes of the droplet dynamics,
we plot in Fig.\ \ref{fig:4} the fixed points in ($\phi_1, \phi_2$) space for increasing Marangoni number $M$
at $\phi_{eq}=0.5$. For the cases $M=3$ and $M=4.5$ the full trajectories are shown. Note that this illustration is a projection onto only two modes of infinitely many modes that make up the full phase space of $\phi$. Starting with the resting state, one has a stable fixed point at $\phi_1=0, \phi_2=0$ for $M\leq 2.7$. Via a subcritical bifurcation the droplet enters the swimming state at the critical Marangoni number $M=2.8$. 
Figure\ \ref{fig:4} demonstrates that for the chosen initial condition $(\phi_1,\phi_2) \approx (0,0)$ both modes $\phi_1$ and $\phi_2$ develop non-zero amplitudes at the same critical Marangoni number. The trajectory in the swimming state does increase its size with increasing $M$, whereas the swimming 
speed decreases until the droplet reaches the stopping state at $M=4.3$. As already observed in Fig.\ \ref{fig:1}, the second mode $\phi_2$, which is symmetric around $\theta=\pi/2$, clearly dominates in the stopping state. In the oscillating regime above $M=9.5$ a stationary solution does not exist. Instead, the dynamics follows a stable limit cycle as the inset (a) in Fig.\ \ref{fig:4} demonstrates for $M=10.5$. 
Finally we remark, since the bifurcation is subcritical, the critical Marangoni number for the onset of the swimming state depends on the chosen initial condition. 
For example, starting the numerical solution at $\phi_1=\phi_2=-0.1$ the critical Marangoni number is $M=1.7$.

\subsection{The active droplet as pusher}

\begin{figure}
\centering
\includegraphics[width=0.3\textwidth, angle=-90, trim=0mm 0mm 0mm 0mm, clip]{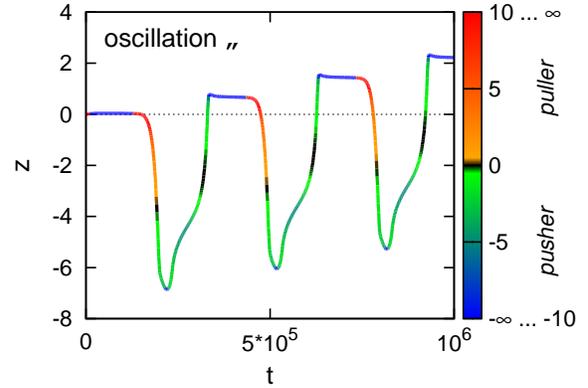}
\vspace{-0.2cm}
\caption{Displacement of oscillating droplet plotted versus time. Color of line shows the value of stirring
 parameter $\beta$.}
\label{fig:5}
\end{figure}

To describe the basic features of a squirming swimmer, it is sufficient to only study the first two modes 
of its surface velocity field \cite{lighthill1952,blake1971,lauga2009,downton2009, zhu2012}.
While the the first mode $\phi_1$ determines the swimming velocity, the dimensionless 
'stirring' parameter $\beta=-\phi_2/|\phi_1|$ characterizes the swimmer type. When $\beta$ is negative, the flow around the droplet is similar to the flow around a swimming bacterium such as \textit{E. coli}. Such a swimmer is called a 'pusher' since it pushes fluid away from itself at the front and at the back. Accordingly, a swimmer with $\beta>0$ is called a 'puller'. The algae \textit{Chlamydomonas} is an example for a puller since it swims by pulling liquid towards itself at the front and at the back\ \cite{lauga2012}. For $\beta\rightarrow\pm\infty$ the droplet becomes a 'shaker', i.e., a droplet that shakes the adjacent fluid but does not swim. If $\beta=0$, the first mode dominates and propels the droplet, as is the case for \textit{Volvox} algae\ \cite{lauga2012}. The classification of the
swimmers according to the `stirring' parameter $\beta$ is illustrated in the inset (b) of Fig.\ \ref{fig:4}. Hydrodynamic interactions between swimmers and with bounding walls depend on their type 
(`stirring' parameter $\beta$) and strongly influence their (collective) dynamics \cite{zoettl2012,evans2011}. For instance, adjacent pushers generally tend to align and swim parallel to each other, i.e., show a polar velocity correlation \cite{ishikawa2006,ishikawa2008}. In fact 
this kind of behavior was observed in experiments of our active droplets \cite{thutupalli2011}. It is therefore of great interest to determine $\beta$. The swimming droplet with $\phi_{eq}=0.5$ is a pusher with $\beta$ ranging from $-0.7$ for $M=2.8$ to $-1.5$ for $M=4.2$. Similar values from $\beta=-0.5$ up to $-7$ were observed throughout the whole swimming regime of the droplet. The stopping droplet is always a shaker with $\beta=-\infty$. Since the limit cycle of the oscillating droplet perambulates all four quadrants of the reduced phase space, it oscillates in the swimming direction as well as in $\beta$, i.e., between being a pusher and a puller. This is demonstrated by the droplet displacement plotted versus
time in Fig.\ \ref{fig:5}.

\section{Conclusions}

We have presented a model for an active squirming droplet with a surfactant mixture at its interface that 
drives a Marangoni flow and thereby drags the droplet forward. Based on a free energy functional for the mixture, we derived a  diffusion-advection-reaction equation for the mixture order parameter at the droplet interface. Relevant parameters
are the Marangoni number $M$ and the reduced relaxation time $\kappa^{-1}$ with which the mixture approaches its equilibrium value $\phi_{eq}$ by bromination or de- and absorption of the surfactants from the surrounding.

As predicted from linear stability analysis in \cite{thutupalli2011}, numerical solutions of the diffusion-advection-reaction equation show that above a critical Marangoni number the resting state of the droplet becomes unstable. The order
parameter develops a non-uniform profile  and the droplet moves with a constant swimming velocity. This only occurs
when the relaxation time $\kappa^{-1}$ (relative to the diffusion time) is sufficiently large. The 
negative stirring parameter $\beta$ identifies the droplet as a pusher in agreement with polar velocity correlations 
found in experiments \cite{thutupalli2011}. A full parameter study in ($\phi_{eq},M$) space also reveals a stopping
droplet, which is a shaker ($\beta=-\infty$), and an oscillating droplet that oscillates between being a puller and a pusher.
We hope that our work initiates further research on the active droplet which constitutes an attractive
realization of the model swimmer called squirmer.

\acknowledgments
We thank S. Herminghaus, U. Thiele, S. Thutupalli and A. Z{\"o}ttl for helpful discussions and the Deutsche Forschungsgemeinschaft
for financial support through the Research Training Group GRK 1558.

\bibliographystyle{eplbib_no_urls}
\bibliography{library}

\end{document}